# Visible Wavelength Division Multiplex System for use as a Instructional Lab System for Higher Education

U. H. P. Fischer (member IEEE), M. Schmidt, T. Volmer, B. Weigl, J.-U. Just

*Harz University of Applied Sciences, Friedrichstr. 57, 38855 Wernigerode, Germany,
phone : +49-(0)3943-659340 fax : +49-(0)3943-659399
Email : ufischerhirchert@hs-harz.de*

Keywords: Wavelength division multiplex, education in optical communications systems, WDM over POF, POF communications systems, polymeric fiber systems

**Abstract:**

The demand for high-speed digital communication such as data, video, and the broadband Internet increases, the required throughput of the modules in communications systems will also increase. In this paper we present an instruction system, which works on the basis of a wavelength division multiplex (WDM) system in the visible spectrum. It is specialised for the academic training at universities to demonstrate the principles of the WDM techniques. It works platform independent in combination with active modules in the training description, short inline videos and interactive diagrams. The system consists of LEDs in different wavelengths using analog and digital signals.

## 1. Introduction

The demand for high-speed digital communication such as data, video, and the broadband Internet increases, the required throughput of the modules in communications systems will also increase[1]. Fast transmitter and receiver modules are basic elements of these systems, which should be able to transmit terabits/s of information via the fiber. Such technologies in turn rely strongly on advanced opto-electronically technologies, and the progress made in optical multiplexing current transmission systems. Time division multiplex (TDM) and wavelength division multiplex[1,2] (WDM) have shown to be the most powerful transmission extension techniques for long-haul in the last decade. The challenge for Universities and technical colleges is to educate technicians and engineers in this new optical communications techniques, especially in WDM applications. In the last time transmission via polymeric fibers (POF) became standard in the automotive industry[3] and in local indoor networks. The combination of WDM with POF will broaden the horizon of low cost optical customer premises networks[4].

In this paper we present an instruction system, which works on the basis of an WDM system in the visible spectrum. It is specialised in the training of technicians in the further education to demonstrate the fundamental principles of data communication via optical fibers. Furtheron, we extended the basic system for using in academic training at Universities. Herein, the most important issues of the WDM optical communication techniques are implemented into an instructional system, which works PC platform independent in combination with active modules in the traning description, short inline videos and interactive diagrams.

---

## 2. Didactical concept and technological design

In this paper we want to introduce a new training WDM system in the visible optical range. It has the intention to be used as an instructional system for in two training stages with ascending complexity. The system is focused on training technicians and students in the 1st *Basic* level and furthermore for higher students at Universities in the 2nd *Advanced* level. For both systems it is planned to develop an interactive education software. With these software it is possible to aid the learning process. At experiment begin the basic knowledge of the students/technicians will be inquired. Therefore it is planned to create a database with questions for the experiment with variations. Furthermore the experiment and the preparation of the protocoll can be performed fully electronically. Besides it is with the help of the software possible to create interactive diagrams to visualize and control the measurement results. Thus, the technicians/students get an overview over the topics and have several tasks to solve, depending to the system outline.

*1st Basic system: training of skilled employees and technicians*
The *Basic* system is focused on technicians in the further education. Its structure is most simply, because it uses WDM structure with very inexpensive equipment. As transmitters simple LED´s in the visible optical domain are used. This simple technology guarantees an inexpensive structure and promotes an intuitive understanding of the WDM technology by the use of signal sources in the visible range, because humans can recognize wavelengths within the range from 450nm - 700nm directly by the eye. Modulating the transmitters is performed by varying the current of the LEDs directly by amplitude modulation (AM). During the lab only three LED sources ( red/660nm, green/550nm, blue/470nm) are used, which are supplemented with a fourth LED which acts as an interference source. A general sketch of the planned structure of the Basic/Extended instructional system is depicted in fig.1.
The optical signals are combined by a star coupler and directly modulated via the bias current with a video signal by approx. 10MHz. The bias offset and the signal strength are the parameters which can be varied at the LEDs at the driver circuit. The individual transmitters must be levelled for the WDM transmission in their transmitting power within 1dB, which is to be realized with by an optical power meter. The three video pictures are sampled with video cameras and represented on video monitors. At the receiver Si photodiodes are used with bandwidths of 10MHz in combination with a transimpedance amplifier. The electrical signals can be varied in offset and strength. The degradation in signal to noise ratio can be evaluated with a scope by variing the length of the POF cable (1-100m).
   Additionally, loss measurements and coupling efficiency of POF fibers (cut and polish, plug connection) in combination with lateral, longitudinal and angle disalignment can be performed by implementing a micrometer stage.
   Tasks for the Basic system:
- PI-curve of several transmitters
- attenuation of different fiber lengths and wavelength
- bandwidth and S-parameter
- influence of EM-fields
- influences of disalignments with the help of a micrometer stage
- signal quality of a video transmission over several fiber lengths and the influence of offset and amplification
- WDM transmission with video signals

*2nd Advanced system: training of university students*
The *Advanced* system is focused on students education at universities. In this module the digital transmission is added to the analog part. Because of the low modulation bandwidth of LEDs, Fabry-Perot diode lasers (LD) are used. The optical signals of the three LDs are combined by a star coupler and directly modulated via the bias current by a bit error tester (BERT) to 155Mbit/s and digital amplitude modulation (NRZ, ASK/PCM). With this set-up all laboratory exercises for optical WDM-transmission can be performed as influence of attenuation, dispersion, optical bandwith, and wavelength shift to any type of transmitted data. In combination with the self developed BERT the eye digramm measurement and bit

**ETOP058 - Visible Wavelength Division Multiplex System for use as a Instructional Lab System for Higher Education**" session "**3 : University level education**", Marseille 2005

error rate tests can be performed. In POF the mode dispersion is the relevant dispersion type which limits the transmission length. The influence of the mode dispersion must be analysed both by experiment and by simulation using standard software PHOTOSS[5]. Spectroscopical detection of the emission of the LDs, their thermal drift behaviour and the filter characteristics of the MUX/DEMUX can be performed using simple spectrometers with a resolution of 1nm (e.g. Newport model OSM-400).

Tasks for the advanced system:
- PI-curve of several transmitters
- attenuation of different fiber lenghts
- bandwidth and S-parameter
- influence of electro-magnetic-fields (EM-fields)
- influences of disalignments by a micrometer stage
- digital transmission
    - bit error rate in dependency on fiber lenghts
    - testing the signal qulaity with an oscilloscope
- WDM transmission with digital signals

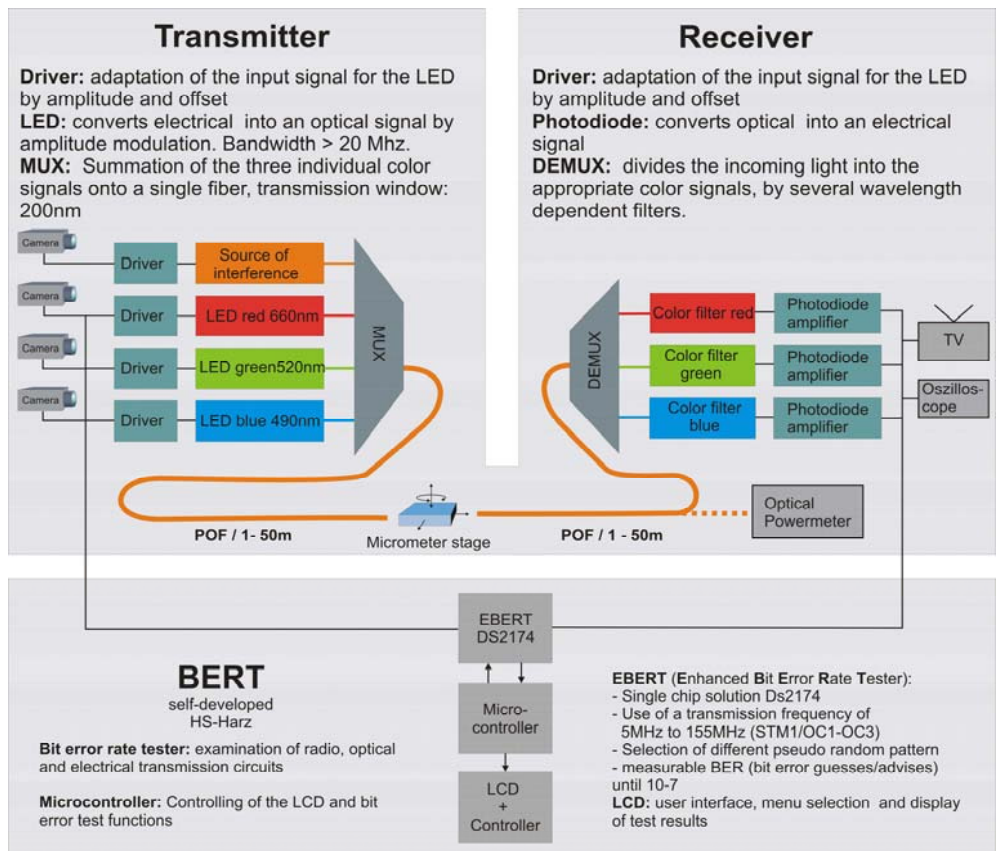

Fig.1 BASIC WDM-lab-training system on the basis of transmission via polymer optical fibers in the visible spectrum

## 3. The transmitter

To convert the electrical into optical signals, which are needed for the transmission, three LEDs (blue@470nm, green@530nm, red@660nm) are used. This represents a cheap solution for the transmitter, because LEDs in this wavelength range are broad available. For analog transmission LEDs with a high linearity are required, to avoid nonlinear distortions.

The input signal will be adapted by a high-ohmic operational amplifier circuit. This voltage amplification is adjustable to adapt different signals to the following circuit. Thus the students/technicians have the possibility to test these influences on the transmission.

---

[5] Simulation software package PHOTOSS: http://www.lenge.de/english/PHOTOSS_overview.php

A DC offset voltage can be applied, which is also adjustable by a high resolution potentiometer. Another part of the transmitter circuit is a voltage-current-converter, which is applied for the modulation of the LED current. The principle is depicted in fig 2. The data to be transmitted is modulating the optical signal.

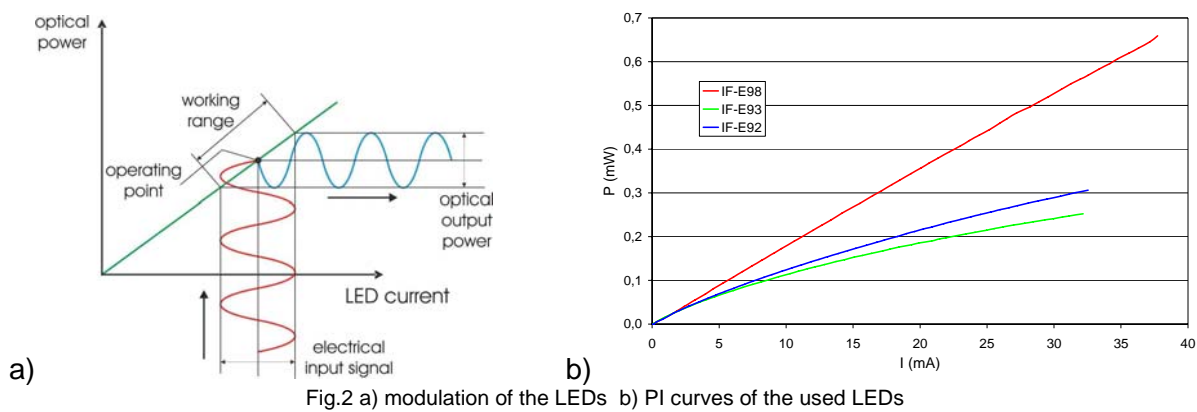

Fig.2 a) modulation of the LEDs  b) PI curves of the used LEDs

This driver circuit is especially adapted to transmit video signals. To generate corresponding input signals there are different possibilities, e.g. a video camera or a test pattern generator can be used. The transmission results can be displayed and judged qualitatively by a monitor on the receiver site. Of course, other types of signals can be transmitted by this system, e.g. a signal generator can be connected and used for transmission of sinus signals. The bandwidth of the transmitter is more than 40MHz for a distance of 20m. With a transmission length of 50m a bandwidth of 35MHz was measured. These results fairly agree with the simulation of the transmission circuit. The bandwidth limitation is resulted by the used operational amplifiers.

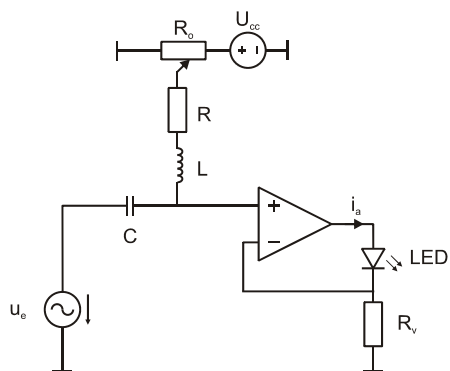

Fig.3 LED driver circuit

Presently transmitters for a digital transmission are tested. These modules will also be integrated in the instruction system.
The transmitter offers a variety of adjustable settings. By working with different parameters the students/technicians can see the influence on the transmission very vivid.

## 4. The Receiver

Receivers are responsible in the teaching system for the transformation of optical signals into electrical signals. The three receivers the multiplexed signals will be seperated into individual signals and adapted to the original signal form. The signal separation into the three original signals red, green and blue takes place via demultiplexing. The DEMUX consist of splitter and chromatic filter. The light is splitted into three rays by two cascading one-to-two toslink[6] splitters. The signal separation takes place via three different chromatic filters, which are directly attached in front of the photodiodes. The receiver technology within the visible range consists of Si pin photodiodes. This kind of photodiode consists of p-type, intrinsic type and n-type semiconductors. The photodiode converts the optical signals by photon-induced pair production into a current. The current is typically in the range of some milliamperes and is transferred into voltage by the following circuit. The appropriate circuit is a two-stage amplifier circuit. The first level is an inverting transimpedance, which is dissipating the current into a voltage. The second level is an inverting amplifier, where amplification and offset are adjustable. Thus the students can change the output signals within certain ranges. The gain can be changed up to 13dB and the offset can be changed from negative and positve voltage. By those changes of the system parameters the students are able to obtain the educational objectives because of changing the signal strength and position by themselves.

---

[6] www.toslink.de

The receivers are developed for analog or digital transmission. For the evaluation of the transmitted signals different devices can be used e.g. network analyzers, oscilloscopes or even television sets or monitors.

## 5. Bit error rate measuring system

Bit error rate testers (BERTs) are predominantly complete units, which can perform a high number of operations. For teaching systems, where the attention is on instructional contents, such devices are only ultilizable with difficulties and very high initial costs. The BERT developed for the teaching system is customized for that use and is the low-budget version of a conventional error rate tester. The developed error rate tester consists of three major components, an error rate tester for error rate measurement, a microcontroller for controlling the measuring device and a display for manual control by the operator. In order to examine a digital transmission circuit, the exit of the error rate measuring device is connected with the transmitter and the entrance is connected with the receiver of the transmission circuit, depicted in fig.4.

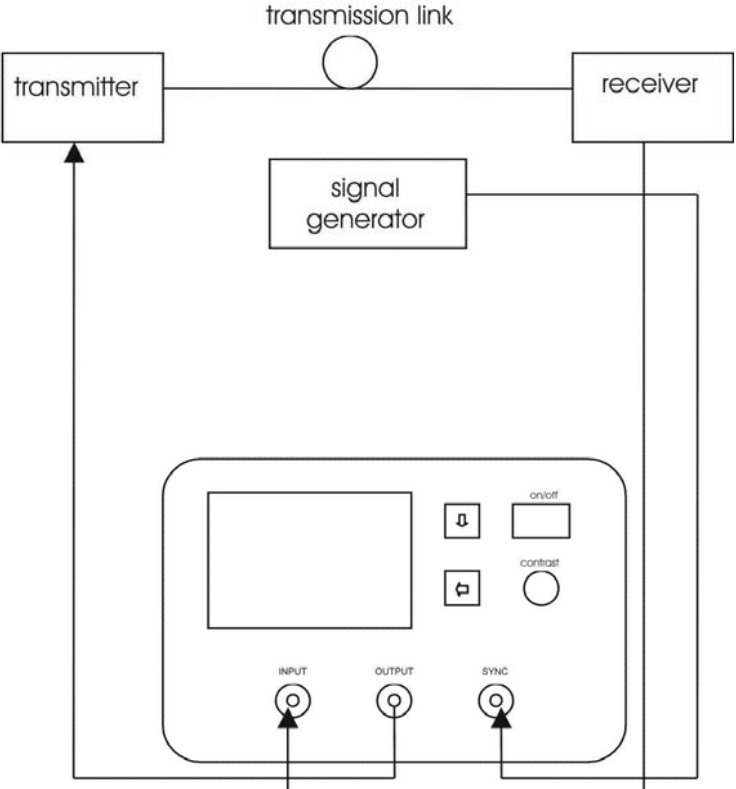

Fig.4 block diagram bit error rate measurement

The transmission frequency used can be either external (max. 155MHz) or internal (40MHz). In order to use an external transmission frequency, a frequency generator has to be attached to the entrance "transmission frequency". The measuring device is controlled via a microcontroller from the 8051 series, the 80C552. This microcontroller also calculates

**ETOP058 - Visible Wavelength Division Multiplex System for use as a Instructional Lab System for Higher Education**" session "**3 : University level education**", Marseille 2005

the error rate after the measurement by the DS2174. The measurement device is operated via a menu on the display. The selection in this menu takes place via a scroll key and an enter key. Before the measuring process can be started, the pattern and the transmission frequency (external/internal) must be selected. The error rate tester (DS2174)[7] allows the selection of coincidental bit-patterns up to a length of 232-1 and self-programmed repetitive patterns up to a length of 512 bytes. In the error rate measuring position, only three patterns can be selected via the display. In the next step, the selected pattern is transferred by means of transmitters on the transmission circuit and can be evaluated at the receiver by the error rate tester. During this evaluation, the sent and received patterns are compared and the bit errors are counted. With this data, the error rate can be calculated by the microcontroller computer 80C552 using Equ. (1)

$$BER = \frac{\text{number of faulty bits}}{\text{number of received bits}} \qquad (1)$$

An important prerequisite for a meaningful measurement is the synchronisation of the output and input port. At the end of each measurement, the result with the pertinent pattern, the number off errors and the BER (Bit error rate) is shown on the display. Bit error rates up to $10^{-7}$ can be measured with this system.

## 6. Instruction System

The first prototype consists of three transmitters and receivers. This system, depicted in fig 3, is able to transmit three analog FBAS[8]-video signals or digital signals up to 10MBit/s. The light of the three transmitters are combined via three conventional mechanically fabricated star couplers 4:1 (DieMount[9]).

To simulate a connector link or a splice, the light is guided via a micrometer stage. Thus, the technichians/students are able to test the influence of lateral, longitudinal and angle disalignments. To separate the signals TOSLINK[10] couplers are used. The insertion loss of this couplers are relatively high.

| Coupler/insertion loss(dB) | 1:2 | 1:4 |
|---|---|---|
| **TOSLINK** | 10,5 $\pm$ 0,9 | 17,8 $\pm$ 0,8 |
| **DieMount** | 5,5 $\pm$ 1 | 9,2 $\pm$ 0,6 |

Table 1 insertion loss of the used couplers

The signal separation is performed by red, green and blue colour filters. The absorption of this filters is relatively high, the absorbtion spectra of the green and red filters is depicted in fig.5. The values are shown in table 2.

---

[7] www.maxim-ic.com
[8] Krisch, L.: „Fersehtechnik: Grundlagen, Verfahren, Systeme", Verlag Vieweg, Braunschweig/Wiesbaden 1993
[9] www.diemount.com
[10] www.toslink.de

**ETOP058 - Visible Wavelength Division Multiplex System for use as a Instructional Lab System for Higher Education**" session  "**3 : University level education**", Marseille 2005

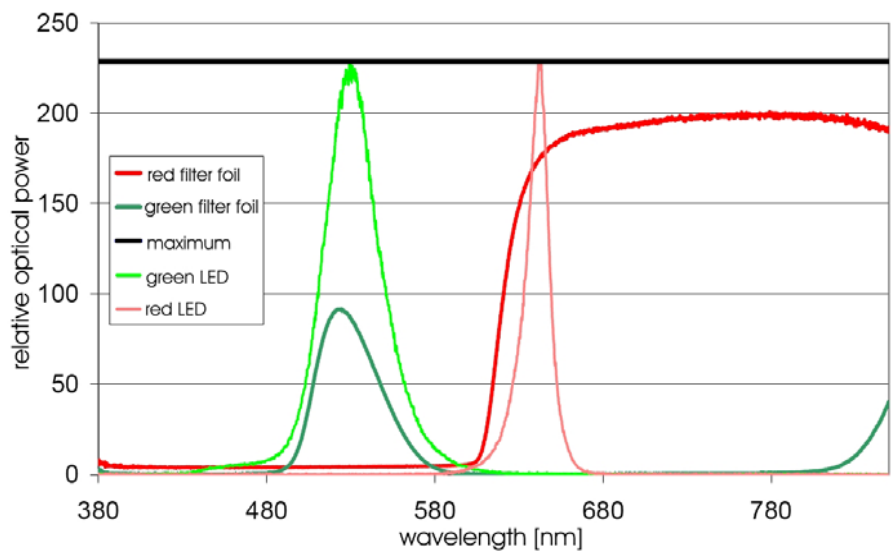

Fig.5 Absorbtion spectra of the red and green filter foil in comparison with the emitted LED power

| Filter foil /insertion loss(dB) | green LED | red LED |
|---|---|---|
| **red filter** | 25,4 | 0,7 |
| **green filter** | 3,5 | 26,8 |

Table 2 Attenuation of red and green filters

This training system prototype was presented at an University campus exhibition in Magdeburg/Germany[11] in July this year (fig.6).
It consists of following parts:
1. video inputs (BNC)
2. regulators for signal amplification, offset and modulation
3. optical outputs (TOSLINK connectors)
4. 1mm SI-POF, pure fiber core without any cladding to make the colours visible
5. DieMount star couplers
6. micrometer stage (x, z and angle)
7. TOSLINK star coupler
8. 1mm SI-POF
9. optical inputs (TOSLINK connectors) with colour filters
10. regulators for amplification and offset
11. electrical outputs (BNC)

---

[11] www.sat2005.magdeburg.de

**ETOP058 - Visible Wavelength Division Multiplex System for use as a Instructional Lab System for Higher Education**" session  "**3 : University level education**", Marseille 2005

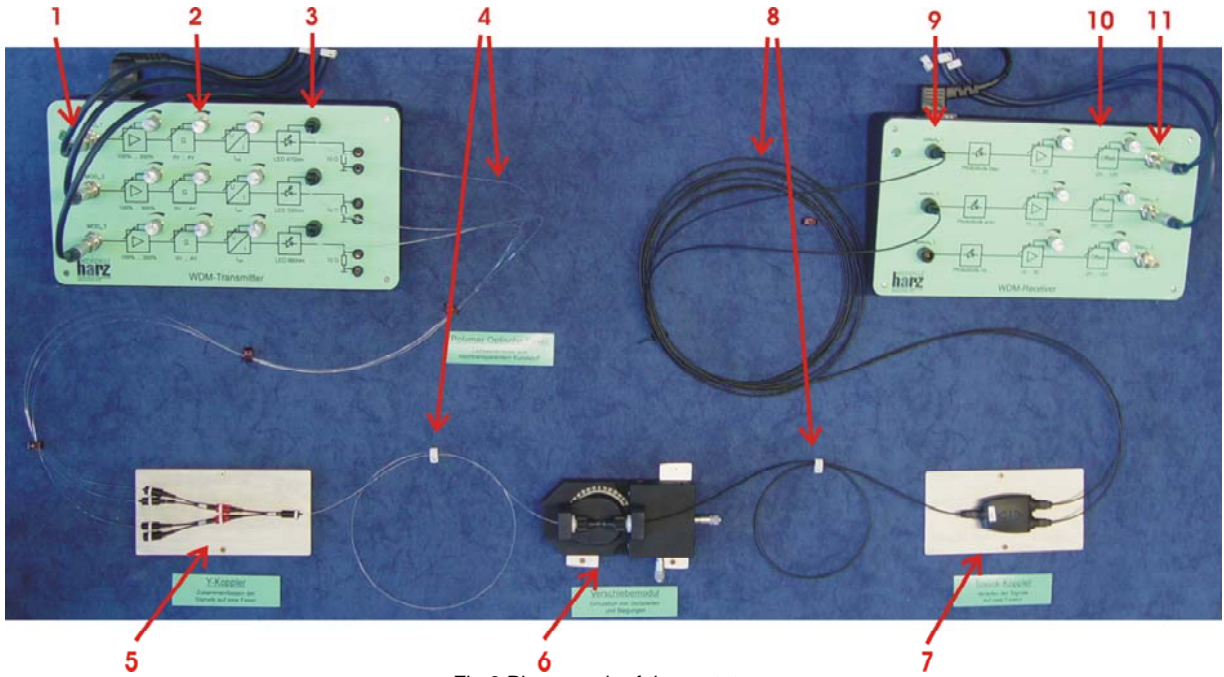

Fig.6 Photograph of the prototype

The bandwidth of the complete system (transmitter, transmission link over 20m and receiver) is about 8MHz, depicted in fig.7. These value is only limited by the receiver. The total attenuation of the system is approximately 38dB.

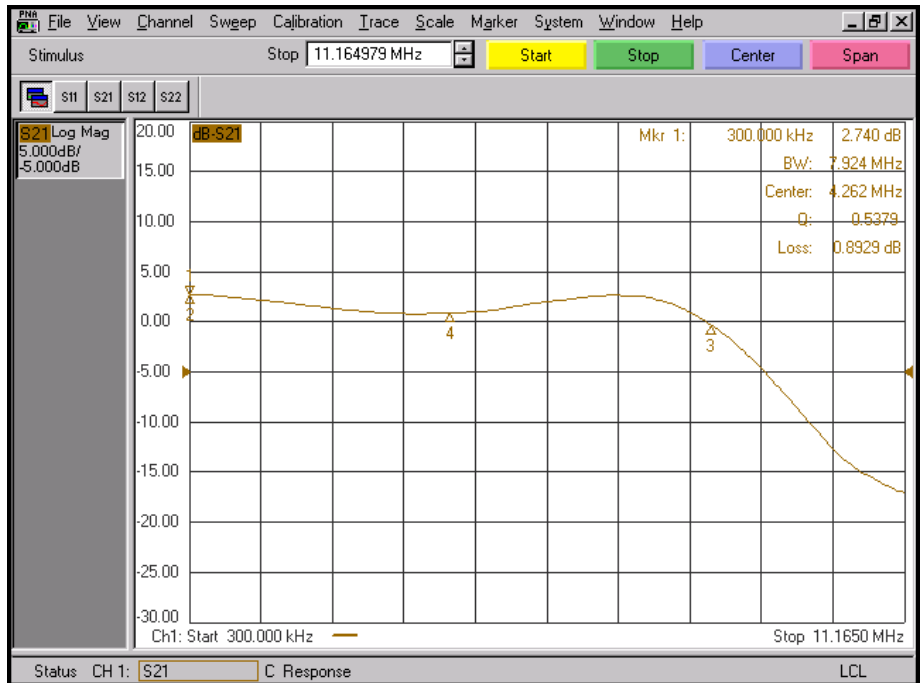

Fig.7 S21 over 20m POF

A further development of single modules for higher modulation frequencies / bit rates are shown in fig.8. For these modules, the circuit and its layout are optimized, respectively. By the use of these modular set-up it is possible to create a modular system for individual applications. The basic system will also consist of three transmitters and receivers with red, green and blue WDM-signals.

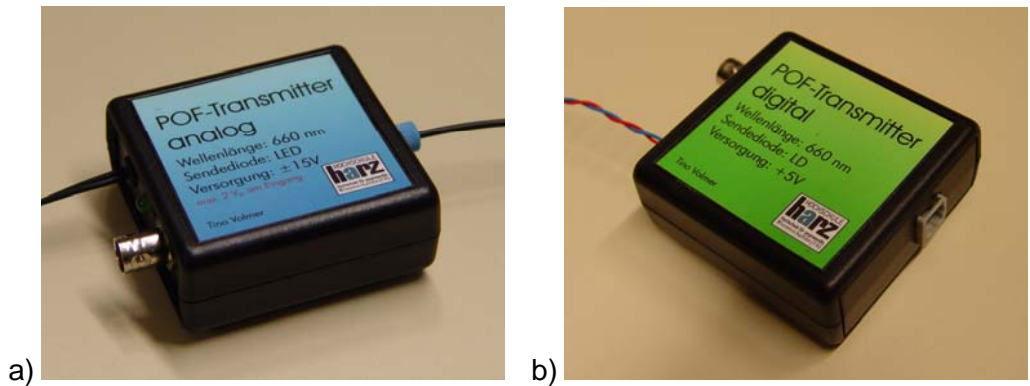

a)  b)

Fig.8 Transmitter modules a) analog, b) digital

Right now a module is developed and tested for transmitting digital signals. It will have the same design like the analog module shown in fig.8. With these new analog transmitter modules bandwidths up to 65MHz are achievable. Thus, it will be possible to combine the transmissions of analog and digital signals via one fiber simultanously.

To reduce the high system attenuation, the next development step will be the design of integrated optical devices for the multiplexing and demultiplexing of the WDM signals. The most important challenge is the development of simple integrated optical MUX/DEMUX (de/multiplexer) for combining/separating the wavelengths using the WDM technology. These devices are designed to transmit up to 8 WDM channels simultaneously. These components are presently patent pending.

## 7. Summary

In this work we present a new WDM training system for technicians and students at universities. The system consists of three LED transmitters in red, blue and green color wavelength. The system can transmit and test either analog video signals or digital signals which are AM or ASK modulated as well as PCM data. With this setup all laboratory exercises for optical WDM-transmission can be performed as influence of attenuation, dispersion, optical bandwith, and wavelength shift to any type of transmitted data. In combination with the self developed BERT the eye diagramm measurement and bit error rate tests can be performed. Spectroscopical detection of the emission of the LEDs and the filter characteristics of the MUX/DEMUX can be performed using simple spectrometers. Additionally, loss measurements of POF fibers (cut and polish, plug connection) in combination with lateral, longitudinal and angle disalignment can be performed.

With this version of a teaching system it is possible to illustrate the optical transmission to students/technicians, especially the WDM technology. The context between emitted power, attenuation etc. can be demonstrated. One of the best advantage of this system is the ability to operate with visible light. Together with the education software this system is good for teaching that makes the subject come alive.

## 8. Acknowledgement

The Department of Research and Development of the Federal Republic of Germany and the Ministry of Sachsen-Anhalt supported this work. We want to thank H. Kragl from DieMount GmbH. for the delivering of the optical couplers and the chromatic filters for the realization of the MUX/DEMUX functionality